The Conference on Pedestrian and Evacuation Dynamics 2014 (PED2014)  14)

# Quantification of bottleneck effects for different types of facilities

Jun Zhang [a,*], Armin Seyfried [a,b]

[a] *Jülich Supercomputing Centre, Forschungszentrum Jülich GmbH, Jülich 52428, Germany*
[b] *Department of Computer Simulation for Fire Safety and Pedestrian Traffic, Bergische Universität Wuppertal, Wuppertal 42285, Germany*

**Abstract**

Restrictions of pedestrian flow could be triggered by directional changes, merging of streams and other changes or disturbances causing effects similar to bottlenecks given by geometrical narrowings. In this contribution we analyze empirically how the types of the changes or disturbances influence the capacity of a facility. For this purpose four types of facilities including a short narrowing, a long narrowing, a corner and a T-junction are investigated. It is found that the reduction of pedestrian flow depends on the shape and the length of the narrowing. The maximum observed flow of the corner (about 1.45 $(m \cdot s)^{-1}$) is the lowest in all facilities studied, whereas that of the short narrowing is highest. The finding indicates that the usage of an unique fundamental diagram for the description of pedestrian flow at different kind of geometrical narrowings is limited.

## 1. Introduction

During the last few decades, research on pedestrian and traffic flow attracted a lot of attention (Appert-Rolland et al. (2009); Bandini et al. (2010); Klingsch et al. (2010); Schadschneider and Seyfried (2009); Schadschneider et al. (2009)). One of the most important characteristics of pedestrian dynamics is the fundamental diagram which states the relationship between pedestrian flow and density. Several researchers have collected information about the relation (Predtechenskii and Milinskii (1978); Weidmann (1993); Fruin (1971); Helbing et al. (2007)). However, the available datasets show surprisingly large differences (Schadschneider and Seyfried (2009)). Facing such problems, several well-controlled pedestrian experiments (Hoogendoorn and Daamen (2005); Kretz et al. (2006a,b); Moussaïd et al. (2009); Liu et al. (2009); Hermes project (2009)) and field studies (Johansson (2009); Young (1999)) have been performed in recent years. Based on the empirical results, several explanations for the discrepancies have been proposed including cultural factors (Chattaraj

* Corresponding author. Tel.: +49-246-161-96554; fax: +49-246-161-6656.
  *E-mail address:* ju.zhang@fz-juelich.de

Peer-review under responsibility of Department of Transport & Planning Faculty of Civil Engineering and Geosciences Delft University of Technology
doi:10.1016/j.trpro.2014.09.008

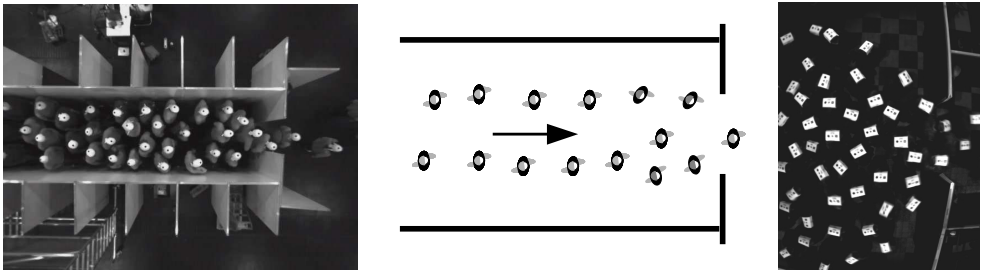

Fig. 1. Sketch and snapshots of pedestrian flow passing through a short narrowing.

et al. (2009)), and differences between unidirectional and multidirectional flow (Navin and Wheeler (1969); Pushkarev and Zupan (1975)).

Pedestrian flow in corridors could be restricted by different factors like narrowings by boundaries, directional changes, merging of streams and so on. Methods in guidelines and handbooks like SFPE (Nelson and Mowrer (2002)), Predtechenskii and Milinskii (1978)), and Weidmann (1993) are mainly based on the assumption that there is a unique density-flow relation for corridors independent from the factors triggering the restriction of the flow. The difference of these factors and the specific properties of different facilities are usually ignored. Whether this assumption is justified or the strength of the restriction depends also on other effects is not well studied.

In this study, we analyze series of well-controlled experiments with restriction of pedestrian flow to quantify the bottleneck effects of different types of facilities including short narrowing, long narrowing, corner and T-junction. The structure of the paper is organized as follows. In section 2 we introduce the setup of different experiments. The bottleneck effect are analyzed in section 3 from the aspects of density, flow profiles as well as the capacities of different geometries. The conclusion will be made in the last section.

## 2. Setup of experiment

In this study, four types of facilities (a short narrowing, a long narrowing, a corner and a T-junction) are considered to investigate their bottleneck effects to pedestrian flow empirically. All data used here are from the experiments performed with up to 400 participants in the wardroom of the Bergische Kaserne Düsseldorf in 2006 and in the fairground Düsseldorf in 2009 in Germany. In these experiments pedestrian movements under congested state were observed and therefore more information on the capacities of different kinds of facilities can be obtained. All of these experiments were recorded by two cameras and the pedestrian trajectories were extracted from the video recordings by the software *PeTrack* automatically (Boltes et al. (2010)). In the following sections, details of the experiments will be described.

### 2.1. Short narrowing

The short narrowing here is a kind of narrowing directly to free space outside (see Fig. 1). Two experiments with this kind of narrowing will be analyzed. One is the unidirectional flow experiment which has been described in Zhang et al. (2011) in great detail. In this experiment, 28 runs were performed in three corridors with the widths of 1.8 *m*, 2.4 *m* and 3.0 *m*. To increase the densities in the corridor, the exits in 9 runs were narrowed which makes the straight corridor becomes a short narrowing (see the snapshot in Fig. 1 left). In these 9 runs the pedestrian flow is restrained by the narrowings and depends on the width of them. The other one is the bottleneck experiment introduced in Rupprecht et al. (2011) and Liddle et al. (2011) which were designed to test the influence of bottleneck length on the flow. Here we only analyze the runs with the bottleneck length of 0.06 *m* which can be regarded as an opening (see the snapshot in Fig. 1 right). The corridor width in front of the narrowing in bottleneck experiment is about 7.0 *m* which is wider than the ones in corridor experiment. Consequently, the pedestrian movements in these two experiments are not really the same, which can be seen from trajectories in Fig. 2. However, this difference has less influence on the flow through the bottleneck according to the findings in Rupprecht et al. (2011).

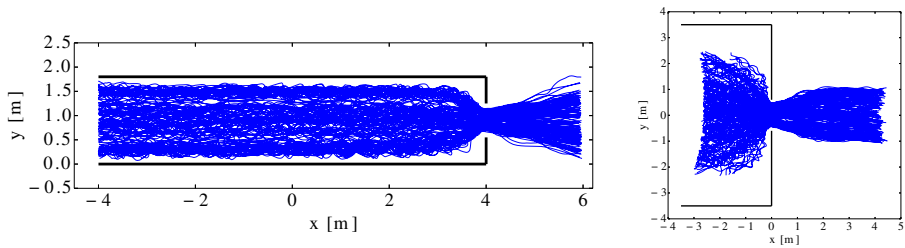

Fig. 2. Pedestrian trajectories from the experiments for pedestrians passing a short narrowing.

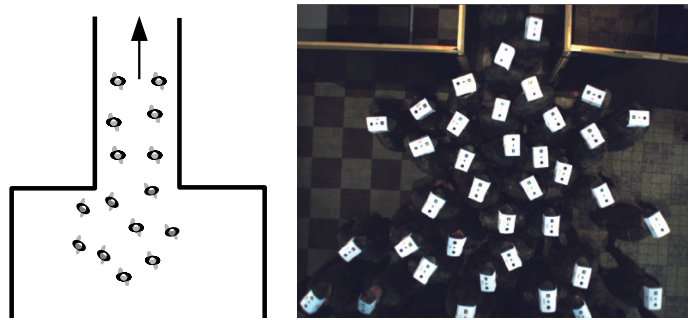

Fig. 3. Sketch and snapshots of pedestrian flow passing through a long narrowing.

### 2.2. Long narrowing

Long narrowing here refers to a geometry formed by two long corridors with different widths (as shown in Figure 3) and pedestrians move from the wide corridor to the narrow one. The pedestrian flow in this kind of geometry is restricted by the capacity of the narrow corridor. For this scenario we analyze the experiment performed in 2006 in the wardroom of the Bergische Kaserne Düsseldorf with a test group of soldiers. The experimental setup allowed the influence of the bottleneck width and length to be probed along with the variable participants. The details on the experiments can be seen in Rupprecht et al. (2011) and Liddle et al. (2011). In this study we only use the data from different bottleneck width but fixed length (4 $m$) and fixed number of participants to investigate the effects of narrow corridor on the flow. Fig. 4 shows the trajectories from two runs of the experiment.

### 2.3. Corner and T-junction

Another configurations considered in this paper are corner and T-junction which can act like a narrowing (Steffen and Seyfried (2009); Yanagisawa et al. (2009)) in certain condition. In the setup of these two experiment, the turning angles of the corners are all 90° and the corridor widths in front and behind the corners are the same for both geometries, which can be seen clearly in Fig. 5 and Fig. 6. Overall two widths 2.4 $m$ and 3.0 $m$ were selected and 36 runs were performed in these two scenarios (see Zhang et al. (2011, 2012a) for details). For the corner scenario, pedestrian streams move from right to left and then turn right. Pedestrian movement in this scenario could be restrained by the turning behavior. Whereas in the T-junction, two streams move towards each other and joined in the junction and form one joined stream. The inflow rates of the two branches are approximately equal and regulated by changing the width of entrances each run. As a result, pedestrian movement in the junction is restrained by the turning and merging behavior as well as the reduction of the effective corridor width.

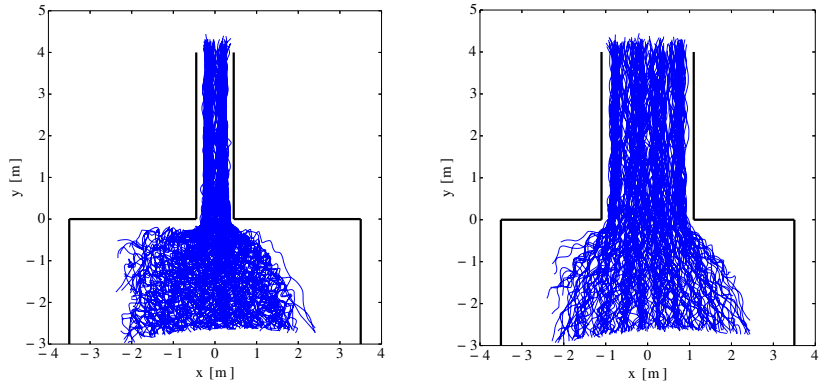

Fig. 4. Pedestrian trajectories from the experiments for pedestrians through a long narrowing.

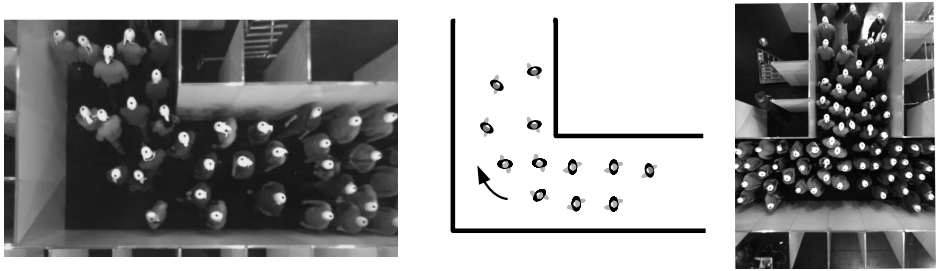

Fig. 5. Sketch and snapshots of pedestrian flow around a corner.

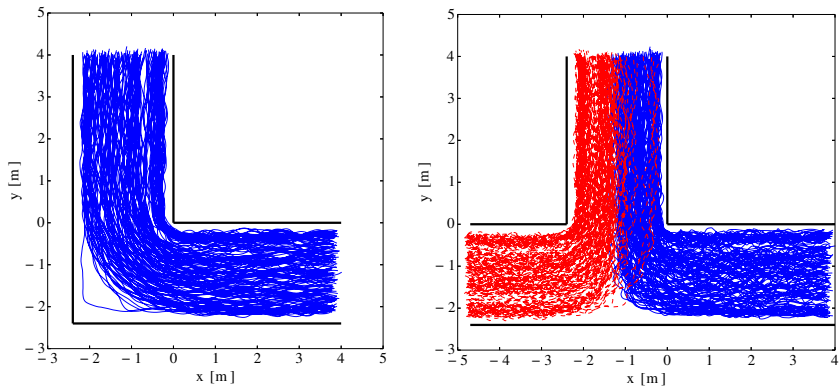

Fig. 6. Pedestrian trajectories from the experiments for pedestrian flows around a corner.

## 3. Analysis and results

In this section, we will investigate the bottleneck effects of the above mentioned facilities from the aspects of spatiotemporal profiles as well as capacities. All the analyses below are based on the trajectories from the experiments.

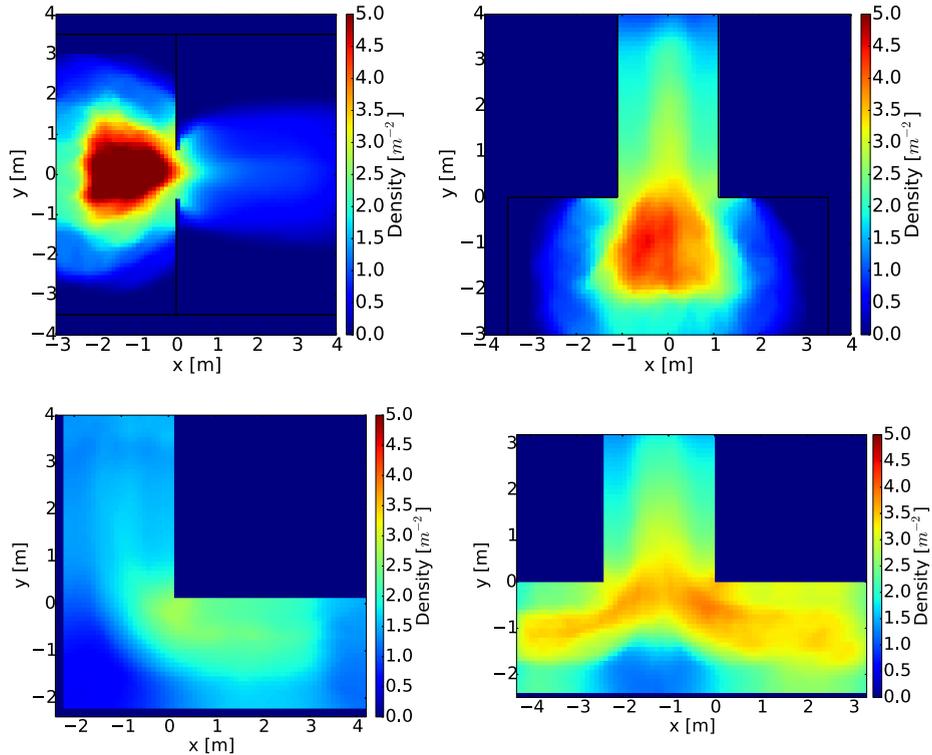

Fig. 7. Density profiles for each scenario in congested state. The maximum of density behind the transition which indicates the supply of pedestrians are different in each scenario.

### 3.1. Spatiotemporal profiles

With the Voronoi method proposed in Steffen and Seyfried (2010) and Zhang et al. (2012a) the measurement area can be smaller than the size of pedestrian. Consequently, the density and velocity integrated over the time can be calculated over small regions (10 cm × 10 cm). To reduce the fluctuations triggered by the start and the end of a run we only consider data in the stationary states to calculate the profiles of density, velocity and flow. These spatiotemporal profiles provide insights into the dynamics of the motion and the sensitivity of the integrated quantities to influencing factors such as boundary conditions.

Considering the particularity of experimental scenarios, the personal space of each pedestrian is limited in a hexagon with the side length of 1 *m* and each Voronoi cell is cut by the hexagon for the short and long narrowing scenarios in the analysis. Fig. 7 shows density profiles for 4 runs of the experiments at congested state. In all cases it is observed that the distributions of the density are not homogeneous over space and transitions from high densities to low densities can be observed. The highest densities for short and long narrowing scenarios appear in front of the bottleneck, whereas in the corner and the T-junction they appear around the inner side of the turning. Behind the transition the density dilutes along with the movement direction in different ways. For the short narrowing case the density decreases the most and becomes homogeneous after moving about 0.5 *m*. In long narrowing and T-junction, higher densities appear in the middle of the corridor due to the boundary effect and the profiles become homogeneous after moving 2 to 3 meters. This different length indicates the restriction effect of bottlenecks along the moving direction. In other word, short narrowing could have least restriction on the flow and have the highest capacity in all these scenarios. Furthermore, for the scenarios with turning (both corner and T-junction) the spaces are not fully utilized even under congested conditions and the place close to the inner side of turning are used more frequently. The actually used corridor width around corner (about 1.6 *m*) is obviously smaller than the whole corridor width (2.4 *m*), which results from the turning behavior and leads to a narrowing of geometry. That means that appearance of a corner could decrease the effective width of corridor and the capacity of a corner would be smaller than common narrowing. This point can be observed from the specific flow profiles in Fig. 8, from where the center and border of the maximal flow can be also obtained. The maximum of specific flow in the four scenarios seems different and changes from about 1.5 to 2.5 $(m \cdot s)^{-1}$. It is worth noticing that the maximal specific flow in the corner is the lowest. This may be due to insufficient inflow (or supply), which is indicated by the small differences of the density in front and behind the corner. In other word, the observed maximum flow even in congested conditions may be not the capacity of the facility and it also depends on the inflow.

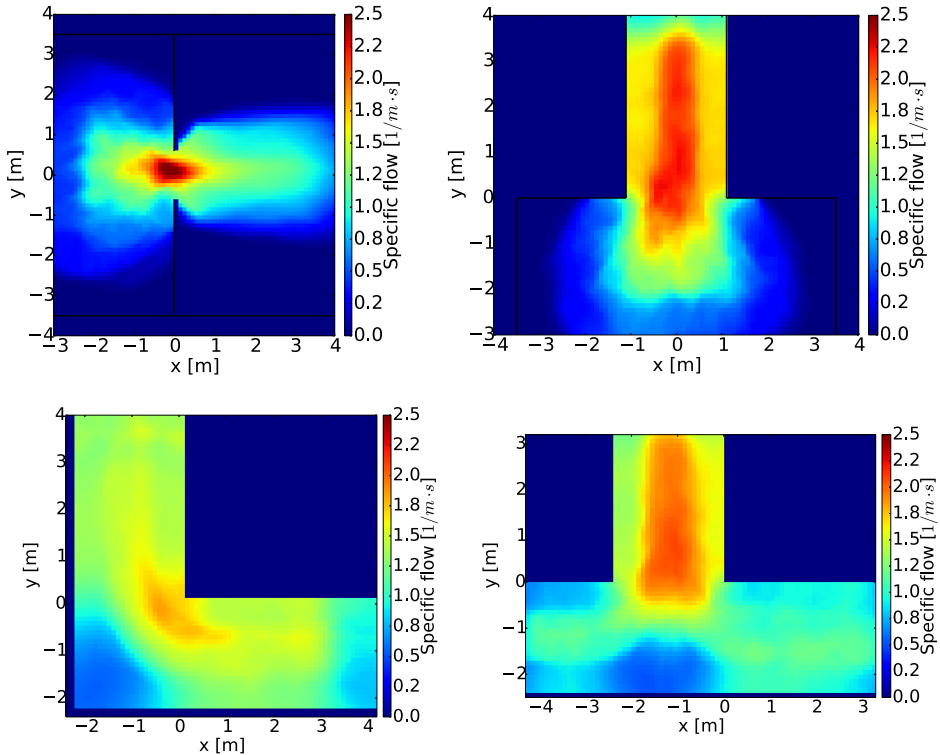

Fig. 8. Specific flow profiles for each scenario in congested state.

### 3.2. Maximum observed flow

In this section, we focus on the maximum flow of these geometries. Normally it is supposed that pedestrian flow in a geometry is determined by its capacity at congested states. Then it would keep a constant when inflow is larger that the outflow. In this situation, congestion occurs in front of the transition and thus the density there would be higher than outside. To obtain the capacities we therefore need select the data under congestion conditions. However, as discussed in section 3.1, the maximum flow in the congested conditions still changes and increase with the increasing inflow. It is still unknown whether the observed maximum flows are the capacities or not.

As discussed before the Method D, Voronoi method, has the advantages of high resolution and small fluctuation of measured densities and velocities. However, the measured densities still depend on the size of the measurement area and the dimension of pedestrian inside of it. In this case, if we use $J_s = \rho \cdot v$ to calculate the specific flow, the measuring error of the density will be introduced and lead to big fluctuations of the results. On the other hand, in the laboratory experiment at the beginning and the end of each run the crowd density is relatively low and the movement is under free flow state. During these periods the outflow is not determined by the capacity. The measured values including these periods could be different (see Rupprecht et al. (2011)) and are not able to reflect the intrinsic property.

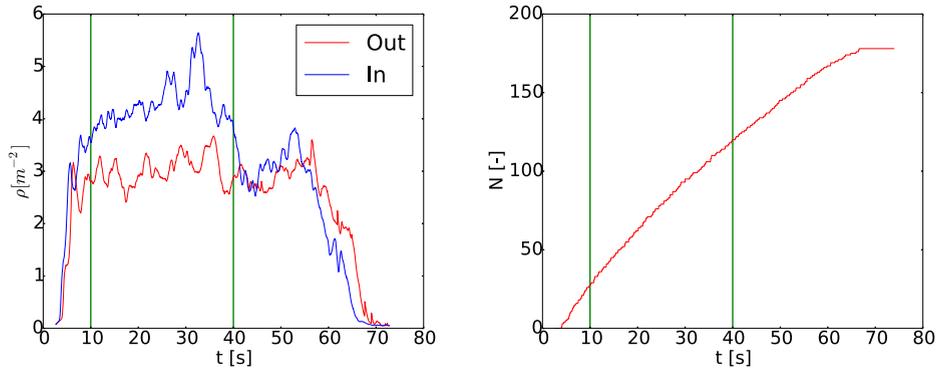

Fig. 9. Selection of the congested state based on the time series of density. The vertical lines show the start and the end of the congested state.

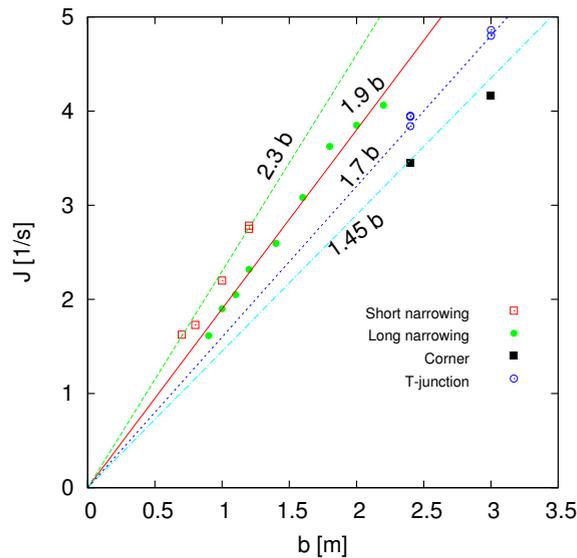

Fig. 10. Capacities for different kinds of geometries. The observed capacities are 2.3, 1.9, 1.7 and 1.45 $(m \cdot s)^{-1}$ in short narrowing, long narrowing, T-junction and corner respectively. Note that the minimum value for corner experiment may results from the insufficient inflow.

Based on the above considerations, here we select the data in congestion state by combining the method A and method D proposed in Zhang et al. (2011) and two steps are adopted to determine the capacities. Firstly we calculate the Voronoi densities in front of $\rho_{in}$ and behind $\rho_{out}$ the transitions or turnings using the same size of the measurement area. Then we choose the periods with $\rho_{in} > \rho_{out}$ for each run from the time series of density (as shown in Fig. 9 left). In the second step, we use method A to count the number of pedestrians passing through the transition line ($\Delta N$) in the selected period $\Delta t$. Then the mean flow rate $J$ can be obtained by $J = \Delta N / \Delta t$ and the results are shown in Fig. 10. Since the specific flow in our experiments is independent on the corridor widths (Zhang et al. (2011, 2012b)), we study the maximum flow of each facility using linear curve fitting method. It is shown that the maximum flow $C$ are not the same for different type of facilities and the order could be $C_{short\ narrowing} > C_{long\ narrowing} > C_{T-junction} > C_{corner}$. The maximum flow for short narrowing is about 2.3 $(m \cdot s)^{-1}$ in the studied experiments, while for long narrowing it is about 1.9 $(m \cdot s)^{-1}$. $C_{corner} = 1.45(m \cdot s)^{-1}$ here is the minimum which may arise from the insufficient supply as discussed in Section 3.1. It is still unknown whether higher values can be obtained for the short narrowing and corner in other condition. But it is obvious that the capacities for a narrowing depend on its length as well as the supply. This result does not agree with most guidelines and handbooks like SFPE (Nelson and Mowrer (2002)), Predtechenskii and Milinskii (1978)), and Weidmann (1993), which base on the assumption that there is a unique density-flow relation for corridors with or without bottlenecks and other narrowing. The finding shows strong limitation of such assumption and raise questions to the methods.

## 4. Summary

In this study, we present series of laboratory pedestrian experiments with restriction of flows in facilities including short narrowings, long narrowings, corner and T-junctions. The whole processes of the experiments were recorded by cameras and the pedestrian trajectories are extracted automatically using *PeTrack*. All analyses in this study are based on these trajectories and only the runs at high densities with congestions are analyzed to investigate the bottleneck effects of different facilities. We quantify the effect from the points of the density and specific flow profiles as well as the capacities for different facilities. It is shown that the restriction effect of a short narrowing on pedestrian flow is smaller than that of long narrowing. In other word, The capacity of a facility depends on the length of its narrowing. Furthermore, the appearance of corner leads to turning behavior which reduces the effective width of the corridor and forms a bottleneck. The different capacities indicates that no unique density-flow relationship can be applied to facilities with different kind of narrowing or geometry. Since the supplies of pedestrians especially at congested station in each scenario are different, further data are still needed to check the capacities of each facility.